%% file: main_arxiv.tex
\documentclass[journal]{IEEEtran}
\IEEEoverridecommandlockouts
\usepackage{graphicx}
\usepackage{etoolbox}
\usepackage{nicematrix}
\usepackage{tikz}
\usepackage{xspace}
\usepackage{bm}
\usepackage{unicode}
\usepackage[flushleft]{threeparttable}
\usepackage{makecell}
\usepackage{enumitem}
\usepackage{mwe}
\usepackage[fontsize=10.5pt]{fontsize}
\usepackage{hyperref}
\usepackage{academicons}
\usepackage[column=0]{cellspace}

\begin{document}

\title{A Dual Convolutional Neural Network Pipeline for Melanoma Diagnostics and Prognostics}

\author{Marie Bø-Sande$^1$*, Edvin Benjaminsen$^1$*, Neel Kanwal$^1$, Saul Fuster$^1$, Helga Hardardottir$^{2,3}$, Ingrid Lundal$^{2,3}$,  Emiel A.M. Janssen$^{2,3}$, Kjersti Engan$^{1,\dagger}$ \href{https://orcid.org/0000-0002-8970-0067}{\includegraphics[scale=0.01]{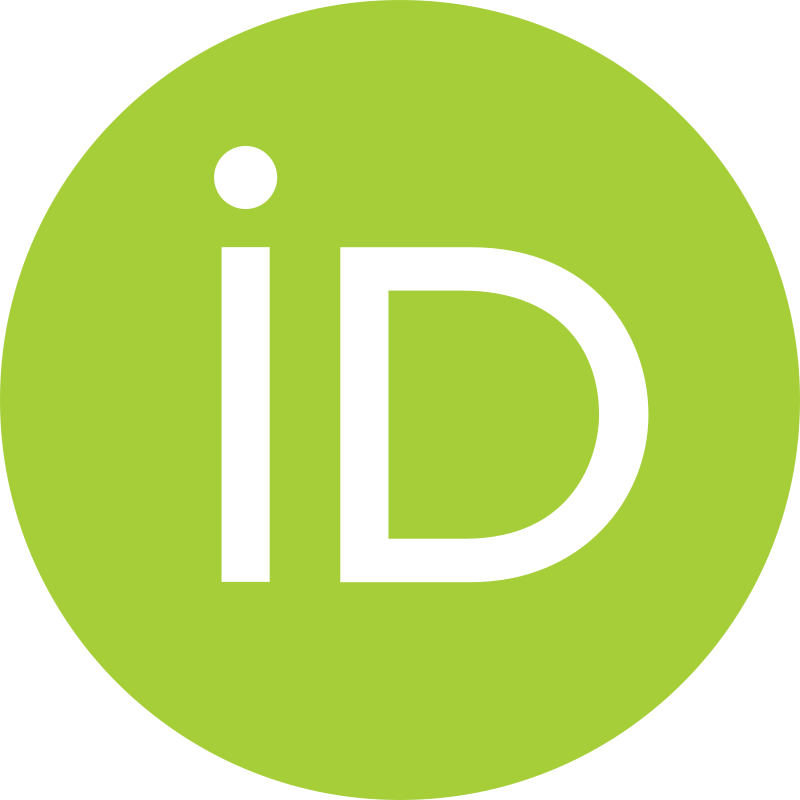}}\\
$^1$ Department of Electrical Engineering and Computer Science, University of Stavanger, Stavanger, Norway\\
$^2$ Department of Chemistry, Bioscience and Environmental Engineering, University of Stavanger, Norway\\
$^3$ Department of Pathology, Stavanger University Hospital,  Stavanger, Norway\\
* These authors contributed equally\\
$\dagger$ Corresponding author: Kjersti.engan@uis.no\\}

\maketitle
\begin{abstract}
Melanoma is a type of cancer that begins in the cells controlling the pigment of the skin, and it is often referred to as the most dangerous skin cancer. Diagnosing melanoma can be time-consuming, and a recent increase in melanoma incidents indicates a growing demand for a more efficient diagnostic process. This paper presents a pipeline for melanoma diagnostics, leveraging two convolutional neural networks, a diagnosis, and a prognosis model. The diagnostic model is responsible for localizing malignant patches across whole slide images and delivering a patient-level diagnosis as malignant or benign. Further, the prognosis model utilizes the diagnostic model's output to provide a patient-level prognosis as good or bad. The full pipeline has an F1 score of 0.79 when tested on data from the same distribution as it was trained on.
\end{abstract}


\input{Introduction}

\input{Data}

\input{Method}

\input{Experiments}

\input{Conclusion}

\input{Compliance}

\input{Acknowledgements}

\bibliographystyle{IEEEtran}
\bibliography{references.bib}

\end{document}

%% file: Introduction.tex
\section{Introduction}
Melanoma cancer is the leading cause of death from skin disease. It begins in the skin cells called melanocytes, and around 30\% starts in existing moles. According to a recent worldwide study, the number of newly diagnosed melanoma cases will rise by more than 50\%, up to 510,000 by 2040, while the number of melanoma deaths will rise by almost 68\%, from 57,000 in 2020 to 96,000 in 2040 \cite{WHO}. In regards to Norway, the annual cancer report shows a 20\% increase in melanoma cases from 2021 to 2022 \cite{CancerNorway}. Coupled with the increased incidence rate, the estimated survival rate for five years following diagnosis varies depending on the stage of melanoma \cite{survivalRate}. Consequently, early detection of melanoma plays a crucial role in the prognostic outcome. 

\begin{figure*}[ht!]
    \centering
    \includegraphics[width=\linewidth]{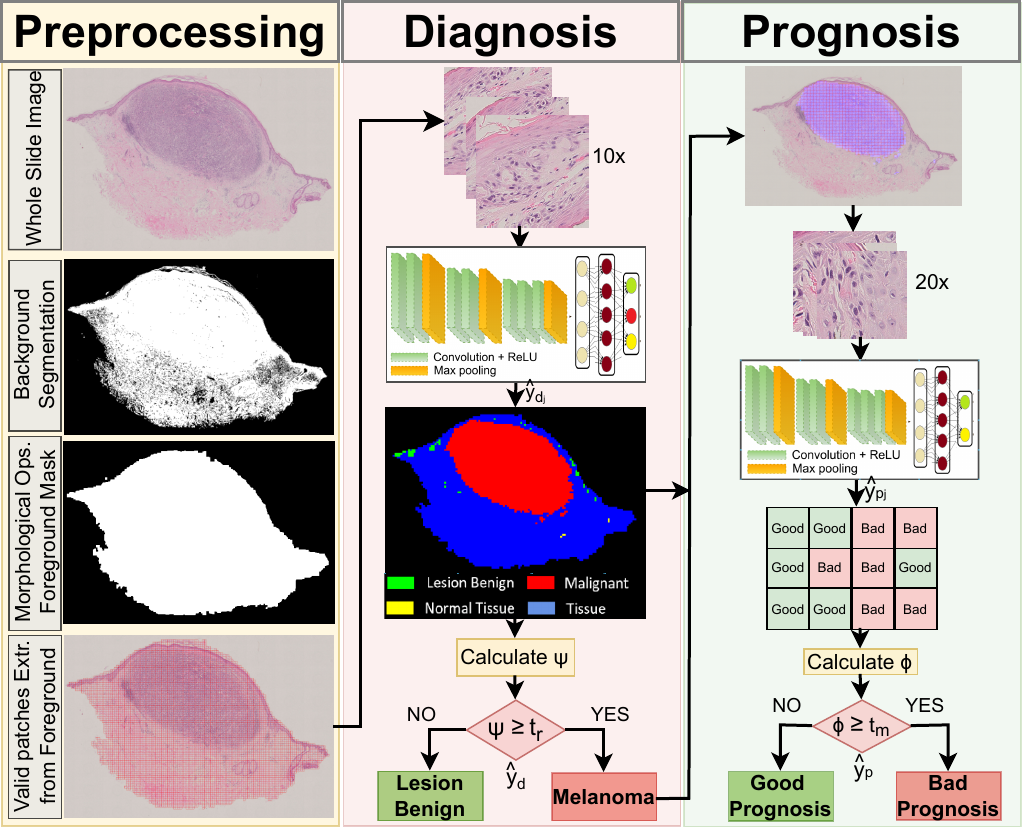}
    \caption{\textbf{An overview of our proposed deep learning pipeline.} \emph{Preprocessing:} Background/foreground segmentation is applied on the whole slide image (WSI) to find tissue regions. Later, morphological operations remove small holes and separate small areas, and then patches are extracted at 10x magnification for the diagnosis model. \emph{Diagnosis:} Prediction for every patch is used to calculate patient-level prediction as benign vs. malignant. \emph{Prognosis:} Detected lesion regions on WSIs predicted as malignant are used to evaluate prognosis at the patient level as good or bad. %The entire pipeline works in an end-to-end fashion.
    }
    \label{fig1}
\end{figure*}

In recent years, digital pathology is becoming mainstream, producing whole slide images (WSIs) as digital microscopy gigapixel images of tissue slides. The process of producing WSIs from biopsies, with its potential artifacts, is described in \cite{Kanwal}. Computational pathology (CPATH) is a growing field dealing with automated solutions for visualization, diagnostics, and prognostics from WSI, using image processing and deep learning (DL).

Melanoma detection using DL techniques has shown promising results, which can help with early diagnosis and treatment decisions ~\cite{Launet2022ASW,kanwal2023detection} \cite{sauter2023deep}. DL algorithms can detect potential regions with melanoma by identifying various cellular and tissue-level features and enhancing diagnostic accuracy~\cite{MLSkinCancer}. %These technological developments may lessen the burden of morbidity and mortality brought on by melanoma and enhance the results of skin cancer treatment.
Some existing methods focus on detecting melanoma by classifying tissue samples and moles either as melanoma or benign nevi \cite{kanwal2023detection, Amor2022MultiResolutionFF, Launet2022ASW}.  Others address the question of prognostic prediction from lesion-tissue of verified melanoma cases, typically with the lesion manually delineated \cite{Andreassen2023DeepLF, forchhammer2022development, Hu2020UsingDL}. Clinical labels are usually patient-based, and providing manually annotated regions for the melanoma tissue is time-consuming. The challenge of having a complete CPATH pipeline that can differentiate between WSIs with benign nevi and melanoma (malignant), segment the melanoma region from a WSI, and make prognostic predictions in the case of melanoma is currently unexplored.
%not discussed in the literature.  

To address this challenge in this work, we are developing a pipeline by integrating two 
%pre-trained 
convolutional neural networks (CNN), as illustrated in Figure~\ref{fig1}. 
%The first model focuses on the detection and localization of melanoma, while the second provides a prognosis for identified melanoma areas. 
By leveraging DL methodologies, the first model of the pipeline identifies melanoma in WSIs and produces a patient-level diagnosis, whereas the second model provides a prognosis for identified melanoma patients. The performance of each CNN model is evaluated individually before integrating them into a comprehensive pipeline.

%% file: Data.tex
\section{Data Materials}
The dataset is collected at Stavanger University Hospital (SUH), Stavanger, Norway. A Hamamatsu Nanozoomer s60 scanner was used to scan a cohort of Hematoxylin and Eosin (H\&E) stained glass slides at 40x magnification, thereafter saved in NDPI format.
Clinical labels give the patient diagnosis as $y_{i}^{d} \in \{0,1\}$ where 1 indicates melanoma and 0 benign nevi, and $i$ is a patient (or WSI) index, and $d$ indicates diagnostic label. For a patient diagnosed with melanoma, the prognostic label was established using follow-up data by considering the occurrence of either local or distant metastasis (bad), or the absence of metastasis (good), within a five-year time frame; $y_{i}^{p} \in \{0,1\}$ where 1 indicates bad prognosis and 0 good prognosis. The dataset is divided into annotated WSIs, $D^{a}$, and not-annotated WSIs, $D^{n/a}$, explained more in the following subsections. 
Later, patches were extracted from the WSIs for analysis. A breakdown of the number of patches extracted from the different sets is shown in Tab.~ref{tab:patches}.

%The pre-trained models are fine-tuned with WSIs provided with rough annotations of regions of interest (ROIs) around lesion areas, along with labels describing the diagnosis of the lesion. 
%The non-annotated dataset used in this work was only provided with patient-based labels which describe the diagnosis of the patient. For a patient diagnosed with malign melanoma, the prognostic label was established by considering the occurrence of either local or distant metastasis (bad), or the absence of metastasis (good), within a five-year time frame.
% Stavanger University Hospital (SUH)

%Annotated dataset - more balanced pga training
%N/annotated dataset - randomly picked out
\begin{table}[h!]
    \centering
      \caption{The amount of patches from each subset of $D^{a}_d$ after patch extraction for the diagnostic model.}
    \label{tab:patches}
    \begin{tabular}{|l|c|c|c|c|}
    \hline
         Label & $D_{train}$ & $D_{val}$ & $D_{test}$ & Total \\
         \hline
         B & 42 420 & 11 660 & 7 190 & 61 270 \\
         M & 215 320 & 25 169 & 37 310 & 27 799 \\
         NE & 2 801 & 657 & 604 &  4 062 \\
         \hline
         Total & 260 541 & 37 486  & 45 104 & 343 131 \\
         \hline
    \end{tabular}
  
\end{table}

\begin{table}[h!]
    \centering
      \caption{The amount of patches of $D^{a}_p$ after patch extraction for the prognostic model.}
    \label{tab:patches}
    \begin{tabular}{|l|c|}
    \hline
         Label & Total \\
         \hline
         Good & 5 542 \\
         Bad & 6 358 \\
         \hline
         Total & 11 900 \\
         \hline
    \end{tabular}
  
\end{table}

\subsection{Annotated WSIs}
%Both models used in this work were trained and validated on WSIs annotated by a pathologist at the hospital. % SUH. 
%ROIs themselves are labeled based on the diagnostic visual content, while the WSI label captures the overall patient diagnosis or prognosis. 
$D^{a}$ is a set of 125 WSIs, 47 benign nevi, and 78 with melanoma. The lesion, or region of interest (ROI), in all WSI, is roughly annotated by a pathologist. The annotated regions of the lesion have two different classes, corresponding to melanoma \textit{M} and benign nevi \textit{B}. In addition, some areas of normal epidermal tissue \textit{NE} are annotated, but not all such areas.  Tissue outside these regions is \emph{not} annotated.  Labels associated with these regions are defined as $y_{ji}^{d}$, where $j$ is a patch index, $i$ is a patient or WSI index. In addition, there are large tissue regions that are \emph{not} annotated in all WSI. 
The diagnosis model used a sub-dataset $D^{a}_d$ of 90 WSIs, where 73 of them were used for training, 8 for validating, and 9 for testing, with approximately 50\% benign nevi and melanoma. The prognostic model used a sub-dataset $D^{a}_p$ of 52 WSIs, all with melanoma, 50\% with bad and 50\% with good prognosis. Some patients were excluded as they were present in both $D^{a}_d$ and $D^{a}_p$. A total of 9 with bad prognosis and 5 with good prognosis. A sub dataset $\hat{D}^{a}_{p}$ was defined comprising all the images from $D^{a}_p$ except for the aforementioned excluded patients, as they were employed for the development of the diagnostic model. There is no overlap from training to validation or test in the pipeline experiments we show.
%, and used stratified k-fold cross-validation to divide the dataset into training and validation in order to avoid overfitting due to the lack of WSIs.
%No WSIs were used to test the performance of the model as they all belong to the same cohort.

\subsection{Non-annotated WSIs}
A dataset $D^{n/a}$ containing 243 WSIs from SUH cohort is provided with patient-level clinical labels without any manual annotation around lesion areas. %All WSIs have been provided with a patient-level label indicating positive and negative diagnostic outcome for benign and malignant, respectively.
Of all 243 WSIs, 110 of them were diagnosed with a benign nevis, and 133 with melanoma, 10 of these with bad prognosis (metastasis within 5 years). The dataset is divided into a train/validation set of 203 WSIs and a test set of 40 WSIs. In the test set, 18 WSIs are labeled as benign and 22 as melanoma, 2 of them having a bad prognosis (metastasis within 5 years).

%Among the patients with melanoma, 12 patients were diagnosed with metastasis prognosis based on a 5-year check-up, while 121 patients had no metastasis prognosis based on a 5-year check-up. Then, the dataset is divided into a validation set of 111 WSIs and a test set of 22 WSIs. In the test set, 20 WSIs are labeled as benign and 2 as malignant.

% A table here would be good

%% file: Method.tex
\section{Method}
% Overall pipeline w. fig

\subsection{Preprocessing}

To enable the analysis of WSIs using CNN, the tissue regions within the WSIs were divided into smaller patches of size 256$\times$256 pixels at different magnification levels (2.5x, 10x, and 40x).  Let  ${\bf x}_{ji}^{10x}$ denote patch $j$ from WSI $i$ at magnification level 10x. The index $i$ denotes the WSI and is sometimes omitted. To separate the tissue from the background, background-foreground segmentation was performed by transforming the RGB images to the HSV color space, and the Hue channel was thresholded within the range of [100-180] to identify purple and pink tones. Morphological opening and closing operations were applied to close holes in the foreground and remove small areas. Grid extraction was applied to extract valid patches as described in \cite{Wetteland2021}.

%Closing and opening morphological operations
%Patches with a 70% overlap between the foreground and annotation masks were extracted, and the "patch-on-fly" technique was utilized for memory-efficient patching.

%To address class imbalance, geometric transformations with random crops of size 224x224 were applied to the underrepresented class.
% Finally, all patches were resized to 224x224 pixels to match the input size of the pre-trained model and normalized using the mean and standard deviation of the training set. 
%This patch-based approach and various preprocessing steps enable the efficient analysis and detection of melanoma in WSIs, facilitating accurate diagnosis and prognosis of melanoma.
% VGG16~\cite{vgg16}
\subsection{Diagnosis}
Valid patches, ${\bf x}_{ji}^{10x}$, from the preprocessing of WSI $i$ are fed into the diagnosis model, providing a patch-level prediction: $f^{d}({\bf x}_{ji}^{10x})=\hat{y}_{ji}^{d}$. The feature extractor of the model is based on the VGG16~\cite{vgg16} architecture with pre-trained weights from ImageNet \cite{deng2009imagenet}. A three-layer classifier of fully connected layers is added. The diagnostic model is fine-tuned using the annotated training data from $D_{d}^{a}$, as in \cite{kanwal2023detection}. The models output layer consists of a softmax giving an array ${\bf v}_{ji}$ of three probability values for each patch, benign nevi (\textit{B}), melanoma (\textbf{M}), and normal epidermal tissue (\textit{NE}). The patch-level diagnosis predictions are denoted as $\hat{y}_{ji}^{d}$, for patch $j$, and WSI $i$. If $max({\bf v}_{ji}) > t_{p}$, $\hat{y}_{ji}^{d}$ is set to the most probable class label (\textit{M, B, NE}). Else, $\hat{y}_{i}^{d}$ is set to  \emph{T} for tissue (i.e., none of the other classes). Thus, we train the model with three labels, but we classify the patches into four classes, $\hat{y}_{ji}^{d} \in \{$\textit{M,B,NE,T}\}. The patient-level prediction $\hat{y}_{i}$ is determined by calculating the ratio ${\psi}$ of number of patches predicted as malignant (i.e. melanoma) over other patches. In this work, two different methods are used to calculate the ratio as shown in Eq.~\eqref{eq:MB_rate_diag} and ~\eqref{eq:MT_rate_diag}. The first ratio, ${\psi}^{MB}$, calculates the ratio between predicted malignant and benign patches, while the second ratio, ${\psi}^{MT}$ calculates the ratio between malignant patches and patches in the entire tissue mask. Let the indicator function, $I(\hat{y}_{ji},\{\textit{M,B}\})=1$ if $\hat{y}_{ji}=\textit{M}$ or \textit{B} and 0 otherwise:

%In \cite{kanwal2023detection}, $t_r$ of $0.04$ in combination with a $t_p$ of $0.999$ was found to be the optimal threshold based on a annotated data set. 

%The dataset provided for this work only consists of WSIs with a patient-level diagnosis and thus can be problematic when not having a true label for each patch. 

\begin{equation}
\psi^{MB}_{i} = \frac{\sum_{j} I(\hat{y}_{ji}^{D},M)}{\sum_{j} I(\hat{y}_{ji}^{D},\{\textit{M,B}\})}  
\label{eq:MB_rate_diag}
\end{equation}

\begin{equation}
\psi^{MT}_{i} = \frac{\sum_{j} I(\hat{y}_{ji}^{D},M)}{\sum_{j} I(\hat{y}_{ji}^{D},\{\textit{M,B,NE,T}\})} 
\label{eq:MT_rate_diag}
\end{equation}

%\begin{equation}
%\psi_{MT} = \frac{\sum_{p} \hat{y}_{pM}}{\sum_{p} (\hat{y}_{pM} + \hat{y}_{pB} + %\hat{y}_{pNT})}
%\label{eq:MT_rate_diag}
%\end{equation}

Thereafter, the ratio is compared with a threshold $t_r$. If ${\psi}_{i}<t_r$ the WSI $i$ is predicted as benign (0); conversely, if ${\psi}_{i}>t_r$ the WSI $i$ is predicted as melanoma (1), i.e. finding patient-level diagnosis label $\hat{y}_{i}^{d}$ as shown in Eq.~\eqref{eq:yp}. 
\begin{equation}
\hat{y}_i^{d} = \begin{cases}
    1, (melanoma) & \text{if } \psi_{i} \geq t_r \\
    0, (benign) & \text{else}
\end{cases}
\label{eq:yp}
\end{equation}

Let $\{{\bf x}_j\}_{M}$ denote the set of patches where $I(\hat{y}_{j}^{d},M) \cap \hat{y}^{d}$, i.e. when the patch is predicted as malignant and the patient is predicted as melanoma, defines the ROI$_{d}$ for further prognostic analysis. 
\subsection{Prognosis}
The prognostic model utilizes a VGG16 backbone, and transfer learning is used with pretrained weights. The classifier of the VGG16 is replaced with three fully-connected layers, the last having softmax activation function, giving a binary output for good or bad prognosis, trained on $D_{p}^{a}$ as in \cite{Andreassen2023DeepLF}. The prognosis model uses malignant patches of the predicted melanoma WSIs for further analysis, i.e. the ROI$_{d}$ defined by $\{{\bf x}_j\}_{M}$ as described in the previous section.  However, the prognostic model operates on a different magnification scale than the diagnostic model.  
%The process therefore involves finding the intersection of the predicted mask with the tissue mask to generate a mask that identifies the specific regions for prognosis prediction. 
Thus, the ROI from the malignant patches is used to extract new patches at 20x magnification with the method of \cite{Wetteland2021}, requiring a minimum 70\% overlap between a valid patch and the ROI.
%alignant patches that have more than 70\% of their area outside the extracted patches are discarded to prioritize meaningful insights for prognosis assessment.This can remove some scattered areas or other outstanding patches. 

The prognosis model provides a patch-level prediction: $f^{p}({\bf x}_{ki}^{20x})=\hat{y}_{ki}^{p}$ for image $i$ and patch $k$ $\in \text{ROI}_{d}$ defined by $\{{\bf x}_j\}_{Mi}$. 
$\hat{y}_{ki}^{p} \in \{0,1\}$ where "1" indicates bad prognosis at patch level.  A threshold $t_{m}$ is used to calculate patient-level $\hat{y}^{p}$ with "1" indicating a bad prognosis and "0" indicating a good prognosis at the WSI level, as shown in Eq.~\eqref{eq:MB_rate_prog}.and Eq.~\eqref{eq:ys}.

\begin{equation}
\phi_{i} = \frac{\sum_{j} \hat{y}_{ji}^{p}}{\sum_{j} (I(\hat{y}_{ji}^{p},0)+ I(\hat{y}_{ji}^{p},1))} 
\label{eq:MB_rate_prog}
\end{equation}

\begin{equation}
\hat{y}^{p}_{i} = \begin{cases}
    1, & \text{if } \phi_{i} \geq t_m \\
    0, & \text{else}
\end{cases}
\label{eq:ys}
\end{equation}
\begin{figure*}[h!]
    \centering
    \includegraphics[width=\linewidth]{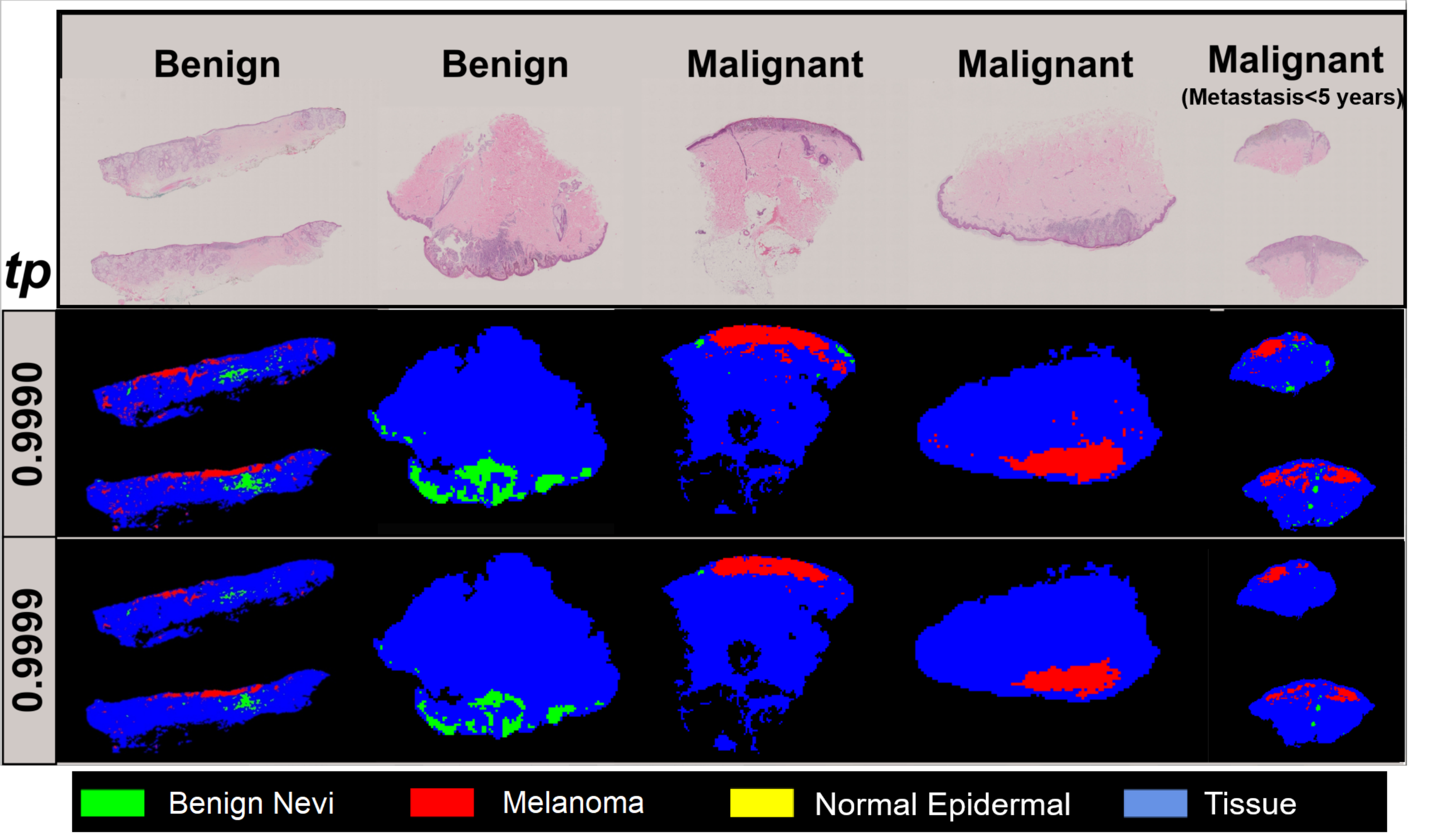}
    \caption{\textbf{Examples of the diagnosis model's patch predictions at different $t_p$ thresholds for $D^{n/a}_{val}$}. Each column displays a WSI with its ground truth label (patient level) at the top. Each row presents the prediction masks for each WSI with the given $t_p$ on the left side.}
    \label{fig:patch predictions}
\end{figure*}

\subsection{Evaluation Metrics}
Let TP, TN, FP, and FN stand for true positive, true negative, false positive, and false negative, respectively.  The accuracy is calculated by $(TP+TN)/(TP+FP+FN+TN)$. Recall/ sensitivity is calculated using $TP/(TP+FN)$. Precision = $TP/(TP+DP)$ and specificity = $TN/TN+FP$. F1 is a weighted harmonic mean of precision and recall. %and is found using $2\cdot Recall\cdot Precision/ (Recall+Precision).$  

%% file: Experiments.tex
\section{Experiments and Results}

Experiments are done by validating the models on the annotated datasets $D^{a}$. The pipeline is tested by comparing prognosis prediction using annotated melanoma ROI as inputs with prognosis prediction using automated ROI, i.e., outputs from the diagnostic model, on the exact same dataset. 
The performance of both models is tested with dataset $D^{n/a}$. We employed 5-fold cross-validation for training our models. We selected the best-performing model and proceeded with it. The results we have shown correspond to the values of the best-performing model trained using cross-validation.
By evaluating the performance of each model separately, it is possible to assess their individual accuracy in addition to testing the combined pipeline.
% Organized
% Try other prognosis patient-level prognosis thresholds.
\subsection{Annotated data}
%Validering, kommenter diagnose, Dp_hatt med og uten annotering

The diagnostic model achieved a performance of 100\% accuracy at the WSI level on the 9 WSIs in the test set of dataset $D^{a}_d$.
%As the dataset $D^{a}_p$ did not contain a sufficient number of WSIs for the application of a test set, 

\subsubsection{Full pipeline test}
The prognosis model's validation process involves comparing the performance of $f^{p}$ when running on a dataset $\hat{D}^{a}_{p}$ with ROI inputs from the manual annotations, ROI$_{a}$, and with ROI from the masks generated by the predictions from the diagnosis model $f^{d}$, ROI$_{d}$. By utilizing the diagnosis model's outputs as inputs for the prognostic model's evaluation, we can effectively assess the performance and accuracy of the prognostic model in a realistic setting. This approach allows us to better understand how the two models work together and how well the prognostic model performs when applied to new, unseen data.

Table \ref{tab:eval_metrics hat_D_p} displays the evaluation metrics after running the prognosis on the dataset ($\hat{D}^{a}_{p}$).

\begin{table}[h]
\caption{Evaluation Metrics after running the prognosis on dataset \(\hat{D}^{a}_{p}\) with annotation masks and generated masks from the diagnosis.}
\centering
\begin{tabular}{|c|c|c|c|c|}
\hline
\textbf{\(\hat{D}^{a}_{p}\)}  & Sens. & Spec. & F1& Accuracy \\
\hline
\hline
ROI$_{a}$ &0.941 & \textbf{0.714} & \textbf{0.821}& \textbf{0.816} \\
\hline
ROI$_{d}$ & \textbf{1.000} & 0.571 & 0.791& 0.763 \\
\hline
\end{tabular}
\label{tab:eval_metrics hat_D_p}
\end{table}
The F1 score and Accuracy are slightly better when using ROI$_{a}$ from manual annotations, but the results look promising for using the automatically found ROI$_{d}$.
 
\subsection{Non-annotated data}

In this experiment, the diagnostic model's performance on the non-annotated dataset $D^{n/a}_{val}$ will be evaluated. The initial thresholds for patch-level classification ($t_p$) and patient-level classification ($t_r$) are set at 0.999 and 0.04, respectively. These thresholds were found in the previous experiment, where the model predicted all images correctly on $D^{a}_d$. The results of the experiment can be found in Table  \ref{tab:eval_metrics diagnosis D_new}.

\begin{table}[h]
\caption{Results from running inference with diagnosis model on $D^{n/a}_{val}$ with thresholds $t_p = 0.999$ and $t_r = 0.04$.}
\centering
\begin{tabular}{|c|c|c|c|c|}
\hline
Eval. Metric & Sens. & Spec. & F1 &Accuracy\\
\hline
Score & 0.977 & 0.146  & 0.728 & 0.601  \\
\hline
\end{tabular}

\label{tab:eval_metrics diagnosis D_new}
\end{table}
The diagnostic model's accuracy is measured to be 0.601, indicating that it correctly classifies only 60\% of the 243 WSIs in dataset $D^{n/a}$. Furthermore, the recall value is calculated to be 0.977, indicating that the model excels at correctly predicting almost all melanoma cases but faces difficulties in accurately predicting benign cases.

\begin{table*}[h!]
%\hspace{-23mm}
\caption{Non-annotated data, pipeline test on the diagnosis $(f^{d})$ and prognostic model $(f^{p})$ on the $D^{n/a}_{test}$. For each of the three tests, performance metrics for the diagnostic model only (D), as well as the complete pipeline (P) is reported as D/P in the metrics.}
\centering
\begin{tabular}{| c| c || c | c | c || c | c | c | c|c|} 
\hline
 Test & Mod & $t_p$ & $t_r$ & $\mathbf{\psi}$ & Spec. &   Sens. &  $F_1$ & Acc \\ [0.5ex] 
\hline
 1 & D/P & 0.9990 & 0.04 & MB & 0.11/0.14  & \textbf{1.00}/1.00  & \textbf{0.73}/0.11  & 0.60/0.18 \\ [1ex]
%\hline
% 1 & P & - & - & - & 0.139  & 1.000 & 0.114  & 0.184 \\ [1ex]
\hline
\hline
 2 & D/P & 0.9990 & 0.04 & MT & \textbf{0.39}/0.07  & 0.82/1.00 &  0.71/0.14  & 0.63/0.14  \\ [1ex]
%\hline
% 2 & P & - & - & - & 0.074  & 1.000 & 0.138 & 0.138  \\ [1ex]
\hline
\hline
 3 & D/P & \textbf{0.9999} & \textbf{0.01} & \textbf{MT} & \textbf{0.39}/0.11  & 0.86/1.00 & 0.73/0.14  & \textbf{0.65}/0.17 \\ [1ex]
%\hline
% 3 & P & - & - & - & 0.107 & 1.000 & 0.138 & 0.167   \\ [1ex]
\hline
\end{tabular}

\label{tab:all_diagnosis_DNewTest}
\end{table*}

The results from this evaluation demonstrate that the diagnosis model shows some level of generalizability for new data with the current settings. However, it is evident that there is room for improvement to enhance its performance further. This highlights the need to focus on parameter tuning and optimization for the model.

\begin{figure}[h]
    \centering
    \includegraphics[width=0.9\linewidth]{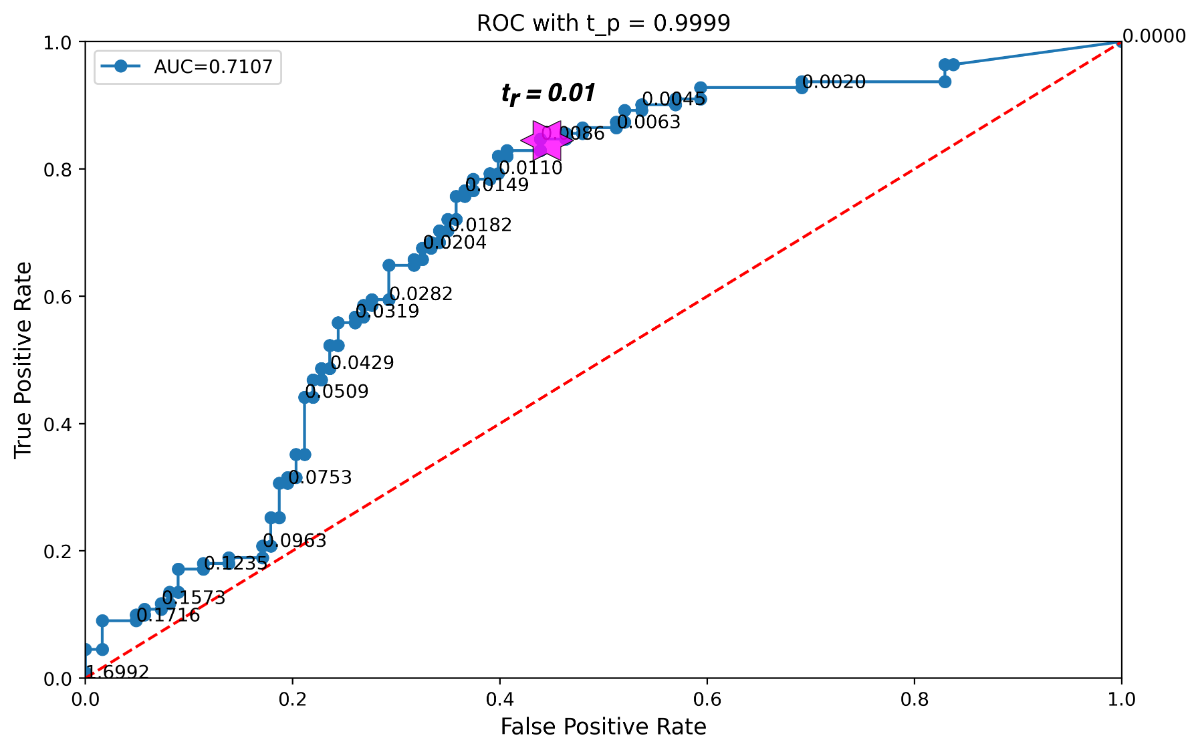}
    \caption{ROC plot for diagnosis model with $t_p=0.9999$ and $MT_{rate}$ method on dataset $D^{n/a}$. }
    \label{fig:roc}
\end{figure}

Dataset $D^{n/a}_{val}$ is used to find optimal threshold $t_p$ and $t_r$. Figure \ref{fig:patch predictions} shows predictions based on different $t_p$. The first column shows an example of a benign slide that is mistaken for melanoma.  There are very few predictions of the \textit{NE} class, probably because the class is underrepresented during training. The distinction between \textit{NE} and \textit{T} has no diagnostic or prognostic relevance, but we keep it as a separate class in case it helps separate benign from malignant cases. In future work, we will investigate this further.

When a lesion is annotated, it is obvious to calculate the ratio of MB patches within the lesion. However, in situations where the lesion size, as well as the tissue size, varies and is unknown, focusing on MT appears to be a more logical approach. For each $t_p$, different $t_r$ was tested with a ROC plot to find the optimal threshold, Figure~\ref{fig:roc} shows the MT ROC plot for $t_{p}=0.9999$ and $t_{r}=0.01$ marked as updated thresholds for the new, larger dataset. For MB, we observed that the optimal threshold choice yields relatively poorer performance across all $t_r$ values. All in all, the increase in AUC score suggests that MT is more effective.

%Ny data, finne nye terskle, ROC
\subsubsection{Non-annotated data pipeline test}
A final test is done using the updated thresholds from the previous experiment, and testing the entire pipeline on the test set from $D^{n/a}$.  
Results are presented in Table~\ref{tab:all_diagnosis_DNewTest}.  The diagnose model performs reasonably well, with F1 scores around 0.73.  The model recall is higher than the specificity, which is also what is desired since it is better to be sure that malignant melanoma is discovered.  The difference between the new and old thresholds is not very large. The prognostic model, however, has a recall of 1 in all experiments and a bad specificity, even if the results in Table~\ref{tab:eval_metrics hat_D_p} are promising.  Prognosis prediction is far more difficult than diagnostic prediction in general, and this shows us that the model has not generalized well enough, and a larger training set with both good and bad prognosis is needed to get general models. In future work, we will investigate using multiple instance learning with a larger dataset for the prognostic part.

%% file: Conclusion.tex
\section{Conclusion}
%This paper presents a CNN pipeline for melanoma diagnosis and prognosis. Our end-to-end pipeline utilizes two state-of-the are convolutional neural network (CNN) models and provides a localization map, along with patient-level diagnostic and prognostic prediction. The By accurately localizing malignant patches in whole slide images (WSIs), our diagnostic model enhances early detection. Coupled with the prognosis model, which leverages the diagnostic output for patient-level predictions, our pipeline offers a comprehensive approach to melanoma assessment for improved patient care. To validate clinical utility, multi-center studies and real-world implementation will be pursued in future, ensuring our pipeline's effectiveness in diverse healthcare settings and populations.
% think about this part as an abstract without an introduction

This paper presents a pipeline putting together two CNN models, one for melanoma detection and localization and one for prognosis prediction on melanoma cases. The pipeline test demonstrates that the prognostic model works similarly well, with F1 scores of 0.82 and 0.79 if the input to the model comes from manual annotation or the output of the diagnostic model when tested on the data set from the same distribution as used to train the models, which is very encouraging. Further updating of the parameters in the diagnostic model showed reasonably good performance for the diagnosis part on a new data set; however, the prognostic model overestimates bad prognosis and should be trained on larger data sets. Prognosis is generally harder to predict.

% \subsection*{Acknowledgement}
% This work has received funding from YYY.

%% file: Compliance.tex
\section*{Compliance with ethical standards}
This study was performed in line with the principles of the Declaration of Helsinki. Approval was granted by the Regional Ethics Committee (No: 2019/747/RekVest). The authors have no relevant financial or non-financial interests to disclose.

%% file: Acknowledgements.tex
\section*{Acknowledgements}
% This research has received fund from research funds.
Thanks to Andres David Mosquera Zamudio for verifying some of the slides. \\
This research has received funding from the European Union's Horizon 2020 research and innovation program under grant agreements 860627 (CLARIFY) and “Pathology Services in the Western Norway Health Region – a centre for applied digitization (PiV)” from a Strategic investment from the Western Norway Health Authority.